\documentclass[aps,pre,reprint,floatfix,amsmath,amssymb,nofootinbib,twocolumn]{revtex4}
\usepackage{graphicx}% Include figure files
\usepackage{dcolumn}% Align table columns on decimal point
\usepackage{bbm}% bold math
\usepackage{epstopdf}
\usepackage[bookmarks=true,colorlinks,linkcolor=blue,urlcolor=blue,citecolor=blue]{hyperref}
\usepackage{color}

\usepackage{setspace} %line spacing
\newcommand{\beq}{\begin{equation}}
\newcommand{\eeq}{\end{equation}}
\newcommand{\beqn}{\begin{eqnarray}}
\newcommand{\eeqn}{\end{eqnarray}}

\usepackage{amsmath}
\usepackage{amssymb}

\begin{document}
%\preprint{arXiv:1105.xxxx}

\title{Numerical optimization using flow equations}

\author{Matthias Punk}
\affiliation{Institute for Theoretical Physics, University of Innsbruck, 6020 Innsbruck, Austria}
\affiliation{Institute for Quantum Optics and Quantum Information, 6020 Innsbruck, Austria}

\date{\today }

\begin{abstract}
We develop a method for multidimensional optimization using flow equations. This method is based on homotopy continuation in combination with a maximum entropy approach. Extrema of the optimizing functional correspond to fixed points of the flow equation. 
While ideas based on Bayesian inference such as the maximum entropy method always depend on a prior probability,
the additional step in our approach is to perform a continuous update of the prior during the homotopy flow. The prior probability thus enters the flow equation only as an initial condition. 
We demonstrate the applicability of this optimization method for two paradigmatic problems in theoretical condensed matter physics: numerical analytic continuation from imaginary to real frequencies and finding (variational) ground-states of frustrated (quantum) Ising models with random or long-range antiferromagnetic interactions. 
\end{abstract}


\maketitle

\section{Introduction}

Optimization problems are ubiquitous in science and engineering and many optimization problems that appear in physics, such as finding the ground state of an Ising spin glass, are NP-hard  \cite{Barahona}. Even though no general algorithms exist to solve such problems in polynomial time, various numerical approaches have been developed in the past to construct approximate solutions, however. 
Here we present an approach to find extremal points of complicated functionals using flow equations. This method combines ideas of homotopy continuation and Bayesian inference. 

Homotopy continuation is a method to solve optimization problems by constructing a one parameter set of solutions which smoothly connect the known extremum of a solvable problem to the extremum of the functional of interest \cite{Allgower}. It has been shown rigorously that this method is able to locate all extremal points for certain problem classes with polynomial nonlinearities \cite{Drexler, Garcia, Wright, Morgan}.
While homotopy continuation methods are only starting to get used in physics \cite{Mehta1, Mehta2, Mehta3}, it is not always obvious to find a solvable reference problem that is suitable as a starting point to construct a homotopy. 
Here we argue that Bayesian inference \cite{Bayes} can be used to define such a solvable reference problem in a very general way, in particular for non-polynomial problems, as long as one is interested in isolated (global) extremal points. The Bayesian approach is based on defining a prior probability, which is updated by new information. In addition we perform a continuous update of the prior during the homotopy flow. This allows to derive a flow equation for the optimization parameters which has fixed points at extremal points of the minimizing functional. Moreover, the Bayesian prior only enters the flow equation as an initial condition.

The paper is outlined as follows: in Sec.~\ref{sec1} we present the flow equation method in general form and apply it to three examples from theoretical condensed matter physics in subsequent sections. In Sec.~\ref{sec2} it is used for numerical analytic continuation, whereas Sec.~\ref{sec3} describes how this approach can be applied to find ground-states of classical (or variational ground-states of quantum) frustrated Ising models. In all three cases the flow equation method gives comparable or better results than standard techniques.

\section{Flow equation method}
\label{sec1}

Here we introduce the flow equation method in its most general functional form and apply it in a discretized version later on.
Suppose we want to find the minimum of an energy functional $E[f(x)]$ with respect to the function $f(x)$ of a real variable $x$.
For the rest of the paper it is crucial that $f(x)$ is positive, which can be achieved in any case after a reparametrization.
A direct solution of the equation $\delta E / \delta f = 0$ determines all the extremal points, but is not practically feasible in most cases. We thus proceed by constructing a flow equation which finds minima of the functional $E[f(x)]$ directly.

Using Bayes' theorem we can recast our minimization problem in a probabilistic form. We wish to maximize the posterior probability $p(f|\mathcal{E}_\text{min})$, i.e.~the probability to find a function $f$ given that it minimizes the energy functional, where $\mathcal{E}_\text{min} = \min_f E[f]$. Quite generally Bayes' theorem expresses the posterior probability to find $f$ conditioned on an arbitrary energy $\mathcal{E}$ as \cite{Bayes}
\begin{equation}
p(f|\mathcal{E}) \sim p(\mathcal{E}| f) \, p(f) \ ,
\label{Bayes}
\end{equation}
where the likelihood function $p(\mathcal{E} | f) = \delta(\mathcal{E} - E[f])$ is the probability to find an energy $\mathcal{E}$ given a function $f$ and we define the Dirac $\delta$ function via
\begin{equation}
p(\mathcal{E} | f) = \lim_{\sigma \to 0} \frac{\exp( -  \big| E[f(x)] - \mathcal{E}  \big| / \sigma)}{2 \sigma}  \ ,
\label{likelihood}
\end{equation}
where $\sigma$ defines the width of the exponential distribution.
Note that this representation of the $\delta$ function is particularly suited for our purpose, but not essential. Lastly, the prior probability $p(f)$ expresses our initial guess for the function $f$. The basic idea of the maximum entropy (MaxEnt) principle is to express the prior probability in the form \cite{Skilling}
\begin{equation}
p(f) \sim \exp \big(\kappa \, S[f(x)] \big) \ ,
\label{prior}
\end{equation}
where
\begin{equation}
S[f(x)] = \int dx  \left( f(x) - f_0(x) - f(x) \, \ln \frac{f(x)}{f_0(x)} \right) 
\label{entropy}
\end{equation}
is a Shannon or von-Neumann entropy and $\kappa$ is an arbitrary coefficient that weighs the relative importance of the prior probability with respect to the likelihood function. Here $f_0(x)$ denotes the prior, i.e.~our initial guess for the solution of the minimization problem. Note that the function $f$ has to be positive, otherwise the entropy is not well defined. The entropic form of the prior probability ensures that no bias is introduced apart from the initial guess $f_0$. 

In the following we are interested in the specific instance where we want to maximize the posterior probability $p(f|\mathcal{E}_\text{min})$. From Eqs.~\eqref{Bayes}, \eqref{likelihood} and \eqref{prior} it is clear
that the the posterior probability is maximized by finding the minimum of the functional
\begin{equation}
Q[f(x);t] = E[f(x)] \, t - S[f(x)] \, (1-t) \ ,
\label{eqQ}
\end{equation}
where we used the specific choice $\kappa = (1-t)/(t \sigma)$ for the coefficient in Eq.~\eqref{prior}.
Again, the parameter $t\in [0,1]$ 
in Eq.~\eqref{eqQ} 
weighs the relative importance of the prior probability with respect to the likelihood function. 
The entropy is maximal for $f(x) = f_0(x)$, which thus corresponds to the minimum of the functional $Q$ at $t=0$. For $t=1$ the minimum of $Q$ corresponds to the minimum of $E$. Starting from Eq.~\eqref{eqQ} we can straightforwardly derive a flow equation for $f(x;t)$ as a function of the homotopy parameter $t$ by requiring that $f(x)$ minimizes the functional $Q[f(x);t]$ for all $t$:
\begin{eqnarray}
0 &=& \frac{\delta Q[f+\delta \! f; t+\delta t]}{\delta \! f(x)} \notag \\
 &=& \frac{\delta Q[f;t]}{\delta \! f(x)} + \int dy \, \frac{\delta^2 Q[f;t]}{\delta \! f(x) \delta \! f(y)} \, \delta \!  f(y) + \frac{\delta \partial_t Q[f;t]}{\delta \! f(x)} \delta t \notag \\
 && + \mathcal{O}(\delta \! f^2, \delta t^2)
\end{eqnarray}
The first term on the r.h.s.~vanishes by construction, since $f$ minimizes $Q$. 
We thus arrive at the flow equation (for notational brevity we write $f(x;t)=f_x$)
\begin{eqnarray}
\frac{\partial f_x}{\partial t} &=& - \int dy \, \left( \frac{1-t}{f_x} \delta(x-y) + t \, \frac{\delta^2 E}{\delta f_x \delta f_y } \right)^{-1}  \notag \\
&& \times \left(\frac{\delta  E}{\delta f_y} - \ln \frac{f_y}{f_0(y)}  \right)
\label{floweq} 
\end{eqnarray}
Note that this flow equation depends explicitly on the prior $f_0(x)$. This explicit dependence can be eliminated if the prior is continuously updated by setting $f_0(x) = f(x;t)$ during the homotopy flow. From this follows that 
\begin{equation}
\frac{\delta S}{\delta \! f(x)} = - \ln \frac{f(x)}{f_0(x)} \Bigg|_{f_0(x) = f(x)} = 0 \ .
\end{equation}
Using a discretized version of Eq.~\eqref{floweq}, which is more suitable for practical applications, the flow equation takes the final form (defining $f_n = f(x_n)$ for some discrete set $\{ x_n \}$)
\begin{equation}
\frac{\partial f_n}{\partial t} = - \sum_m \, \left( \frac{1-t}{f_n} \delta_{m,n} + t \, \frac{\partial^2 E}{\partial f_m \partial f_n } \right)^{-1} \frac{\partial  E}{\partial f_m} \ .
\label{floweqdiscrete}
\end{equation} 
This equation describes the flow of the variables $f_n(t)$ as a function of $t$ towards a fixed point where the energy functional is minimal, i.e.~$\partial E / \partial f_m = 0$ with a positive Hessian matrix $\partial^2 E /(\partial f_n \partial f_m)$. Note that the prior enters only as an initial condition. 

A numerical solution of Eq.~\eqref{floweqdiscrete} is straightforward, and the computationally most costly step is the numerical inversion of the Hessian matrix $\partial^2 Q / \partial f_n \partial f_m$.
Since the majority of interesting minimization problems have many local minima, the success of the flow equation method to find the global minimum of $E[f_n]$ depends on the possibility to choose an initial condition which lies in the basin of attraction of the fixed point corresponding to the global minimum.  
The flow equation \eqref{floweqdiscrete} preserves symmetries during the flow, which makes this method especially suited to find solutions where the global minimum is constrained by symmetries which can be enforced by choosing a proper initial condition (we give an example in Sec.~\ref{sec3b}). This is in contrast to traditional methods based on random updates, such as simulated annealing \cite{Kirkpatrick}. Note that the flow equation also has fixed points at $f_n=0$, which is particularly useful for some applications, as will be shown in Sec.~\ref{sec2}.

The continuous update of the prior can lead to a slower convergence than the standard homotopy continuation without update of the prior for some problems. In fact, it might happen that one does not reach the energy minimum during the flow from $t=0$ to $t=1$. In this case the flow has to be restarted at $t=0$ with the output of the previous run as initial condition until one converges to the minimum. Alternatively it is also possible to rescale the energy term with a large pre-factor, which leads to a faster convergence. By contrast, homotopy continuation guarantees to reach an energy minimum at $t=1$ when using Eq.~\eqref{floweq} without prior update. 
It is important to emphasize, however, that the continuous prior update has practical advantages in many cases. Indeed, when solving the flow equation numerically, a truncation error will be accumulated due to the finite step-size. The advantage of using the flow equation \eqref{floweqdiscrete} is that it is less prone to truncation errors, because the flow is directed towards the fixed point $\partial E/\partial f_n=0$. Without prior update the truncation error can be kept small only by reducing the step-size, which is computationally costly for problems with a large number of variables. The question which version of the flow equation works better thus depends on the problem at hand. 
 
For the problem of numerical analytic continuation to be discussed in Sec.~\ref{sec2}, the continuous update of the prior has an additional conceptual advantage. In this case the flow has to be stopped at some $t<1$, because the Hessian $\partial^2 E /(\partial f_n \partial f_m)$ is singular. For this reason the result depends explicitly on the prior in conventional MaxEnt approaches. By contrast, using \eqref{floweqdiscrete} the dependence on the prior is only implicit through the initial condition.

The possibility to get stuck in a local minimum is inherent to all numerical minimization algorithms and the flow equation approach is not an exception. Compared to standard gradient-based methods this approach is less sensitive to the choice of initial conditions, however, and the probability to find the global minimum is substantially higher for the problems we've studied so far.

Finally, we note that the functional $Q$ in Eq.~\eqref{eqQ} takes the form of a free energy. The flow equation approach can thus be viewed as a deterministic annealing method, where the energetic ground-state is found by following the minimum of the free energy while decreasing temperature \cite{Rose}.

\section{Application: Numerical analytic continuation}
\label{sec2}

Following the derivation of the flow equation \eqref{floweq} above, it is obvious that our starting point is reminiscent of the MaxEnt method which is used routinely to analytically continue imaginary time data to real frequencies \cite{Silver, Jarrell}. Our approach can be viewed as an alternative method to solve the MaxEnt equations numerically.

Numerical analytic continuation can be reduced to the problem of finding the spectral function $A(\omega)$ as function of real frequency $\omega$, given numerical Green's function data $G(i \omega_n)$ on the imaginary (Matsubara) axis. They are related by the Cauchy integral
\begin{equation}
G(i \omega_n) = \int d\omega \, \frac{A(\omega)}{\omega- i \omega_n} \ .
\label{cauchy}
\end{equation}
After discretising the integral, the problem of finding $A(\omega)$ amounts to a matrix inversion problem with the severe complication that the matrix is almost singular: many eigenvalues are very close to zero and thus tiny numerical errors blow up exponentially in the matrix inversion process.
Numerical analytic continuation is thus per se an ill defined mathematical problem. Nevertheless, ideas have been put forward to regularize the matrix inversion without introducing too many artificial features. The most widely used approach to analytically continue noisy Green's function data from Monte Carlo simulations is the MaxEnt method \cite{Silver,Jarrell}, which can be derived using Bayesian inference as outlined above, but other methods have been put forward as well \cite{Sandvik, Prokofev}.

Fitting the spectral function $A(\omega)$ to the Green's function data $G(i\omega_n)$ amounts to minimizing the mean square distance
\begin{equation}
\chi^2 = \frac{1}{N} \sum_{n=1}^N \frac{1}{\sigma_n^2} \left| G(i \omega_n) - \int d\omega \, \frac{A(\omega)}{\omega- i \omega_n}\right|^2 \ ,
\label{chi2}
\end{equation} 
where $\sigma_n$ denotes the standard deviation of the noisy data $G(i \omega_n)$ from the expected value. 
Finding the minimum of $\chi^2$ via $\delta \chi^2 / \delta A = 0$ leads immediately to the ill defined matrix inversion problem in Eq.~\eqref{cauchy}. In order to regularize the inversion MaxEnt adds an entropy term and minimizes $Q[A] = - \kappa \,  S[A] + \chi^2[A]/2$ instead. Again, the entropy term depends on a prior $A^0$. Apart from choosing the prior, the main problem in traditional MaxEnt is to determine the weight parameter $\kappa$ and the standard approach nowadays is to use Bryan's algorithm \cite{Bryan}.

\begin{figure}
\begin{center}
\includegraphics[width=0.9\columnwidth]{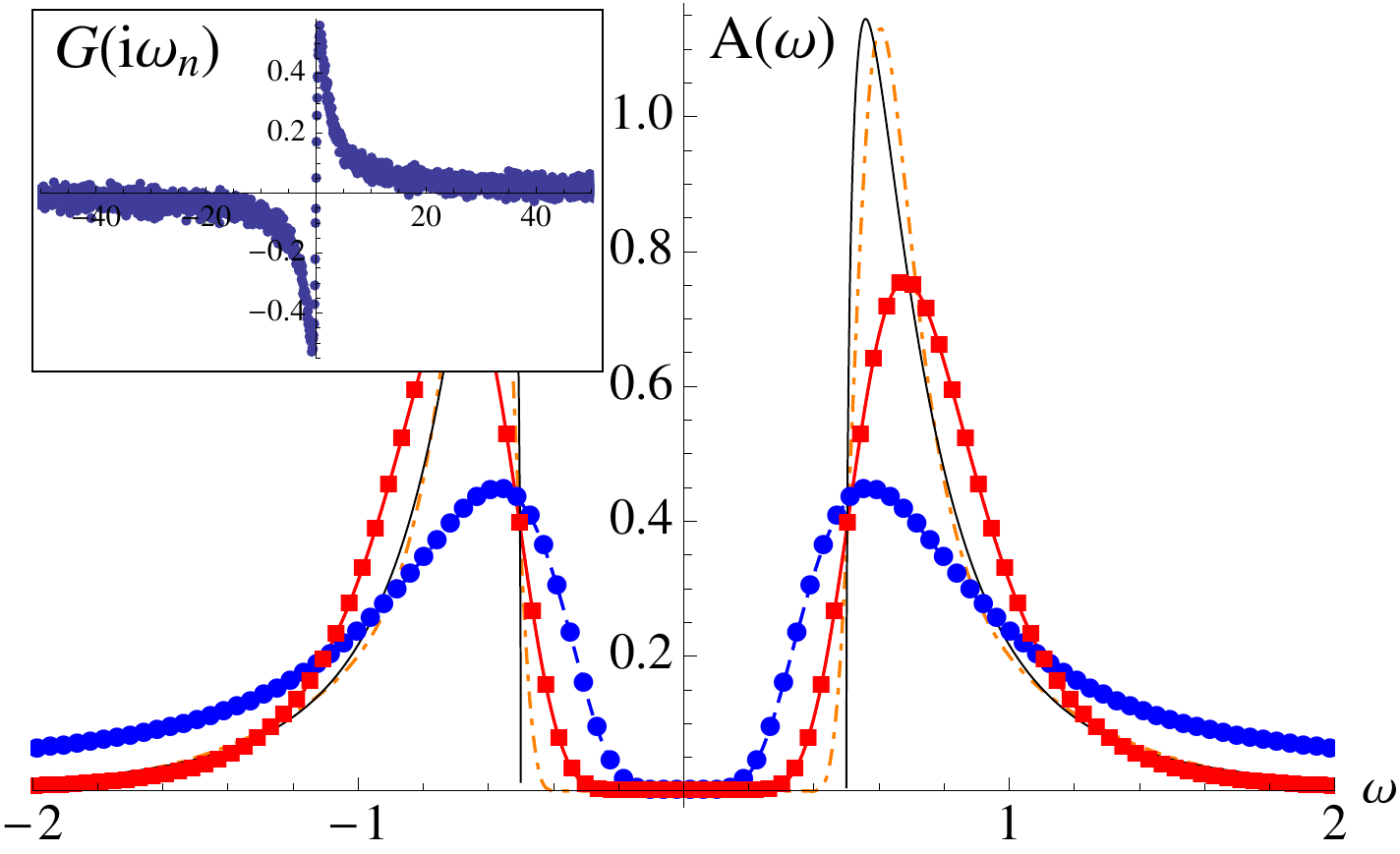}
\caption{(Color online) Applying the flow equation method to perform numerical analytic continuation. Green's function data on the imaginary frequency axis was generated from the model spectral function in Eq.~\eqref{modelA}, shown as black solid line, using Eq.~\eqref{cauchy}. The insert in the top left corner shows the mock data for $G(i \omega_n)$ including statistical noise with $\sigma_n =0.02$. Red line with squares: numerically reconstructed spectral function $A(\omega)$ obtained using the flow equation method. Blue dashed line with dots: historic MaxEnt \cite{Jarrell} result for comparison. The orange dash-dotted line shows the spectral function obtained using the flow equation method for noiseless Green's function data. A flat prior spectrum $A^0(\omega) = const.$ has been used in all cases.}
\label{fig:ac}
\end{center}
\end{figure}

Here we solve the analytic continuation problem by minimizing $\chi^2$ using the flow-equation \eqref{floweqdiscrete} after discretising the integral in \eqref{chi2}. 
Green's function data from Monte Carlo simulations has a statistical error, thus we don't want to follow the flow to the fixed point $\delta \chi^2 / \delta A = 0$, as this amounts to fitting the noise. Moreover, the flow equation is unstable close to $t=1$, where the Hessian is almost singular. Consequently, we stop the flow at some $t$ and we face a similar problem as determining the weight $\kappa$ in the traditional MaxEnt approach.
The naive criterion of stopping the flow when  $\chi^2 \simeq 1$ is similar to the so-called historical MaxEnt scheme \cite{Jarrell}.
For the case of noisy Green's function data we rather choose to stop the flow when $| \nabla_{A} \chi^2|$ is minimal. Indeed, while $\chi^2$ is monotonically decreasing during the flow, the norm of the gradient of $\chi^2$ with respect to $A$ has a minimum at some point for noisy data, if the problem is sufficiently oversampled (i.e. the number of data points $G(i \omega_n)$ is much larger than the number of sampling points form the discretisation of $A(\omega)$). We identify this minimum with the point beyond which one starts to over-fit the noisy data. For clean data without noise we terminate the flow before the inversion of the Hessian becomes numerically unstable. In passing we note that the $\sim \delta_{m,n}/f_n$ term in Eq.~\eqref{floweqdiscrete}, which regularizes the matrix inversion, is precizely the discrete Fisher information metric \cite{Fisher}. 
 
 In order to check the applicability of our method, we perform numerical analytic continuation for a simple model spectral function of the form
 \begin{equation}
 A(\omega)= 4 \, \sqrt{1-\frac{\Delta^2}{\omega^2}} \, \frac{\Delta^4}{\omega^4} \ .
 \label{modelA}
 \end{equation}
We generate Green's function data from this model for $\Delta=0.5$ using Eq.~\eqref{cauchy} and add random Gaussian noise with $\sigma_n=0.02$. The results of the flow equation method are shown in Fig.~\ref{fig:ac} together with results obtained with the historic MaxEnt approach. In all cases the same flat prior spectral function $A^0(\omega) = \text{const}$ has been used. 

Fig.~\ref{fig:ac} shows that the flow equation method gives slightly better results compared to historic MaxEnt. The flow equation is particularly useful when the spectral function is zero in some frequency interval, i.e.~if there is a gap in the spectrum, because Eq.~\eqref{floweqdiscrete} has fixed points at $f_n=0$. By contrast, standard MaxEnt has problems if the spectral function is close to zero, because the entropy term diverges. Note that the additional fixed points at $f_n=0$ are only approached asymptotically and thus it is unlikely that the flow is forced into a poor minimum. Indeed, for $f_n \to 0$ the flow equation simplifies considerably and has the solution $f_n(t) \sim (1-t)^\alpha$ with $\alpha = \partial E/ \partial f_n$.

Moreover, the prior spectrum enters the flow equation only as an initial condition. For this reason we can expect that different priors lead to the same spectral function as long as the initial conditions lie in the basin of attraction of the same fixed point. Consequently, our method might be useful as an alternative to Pad\'e approximants for data without statistical noise \cite{Pade}.

\section{Application: Ground states of (quantum) Ising models}
\label{sec3}

 \begin{figure}
\begin{center}
\includegraphics[width=0.8\columnwidth]{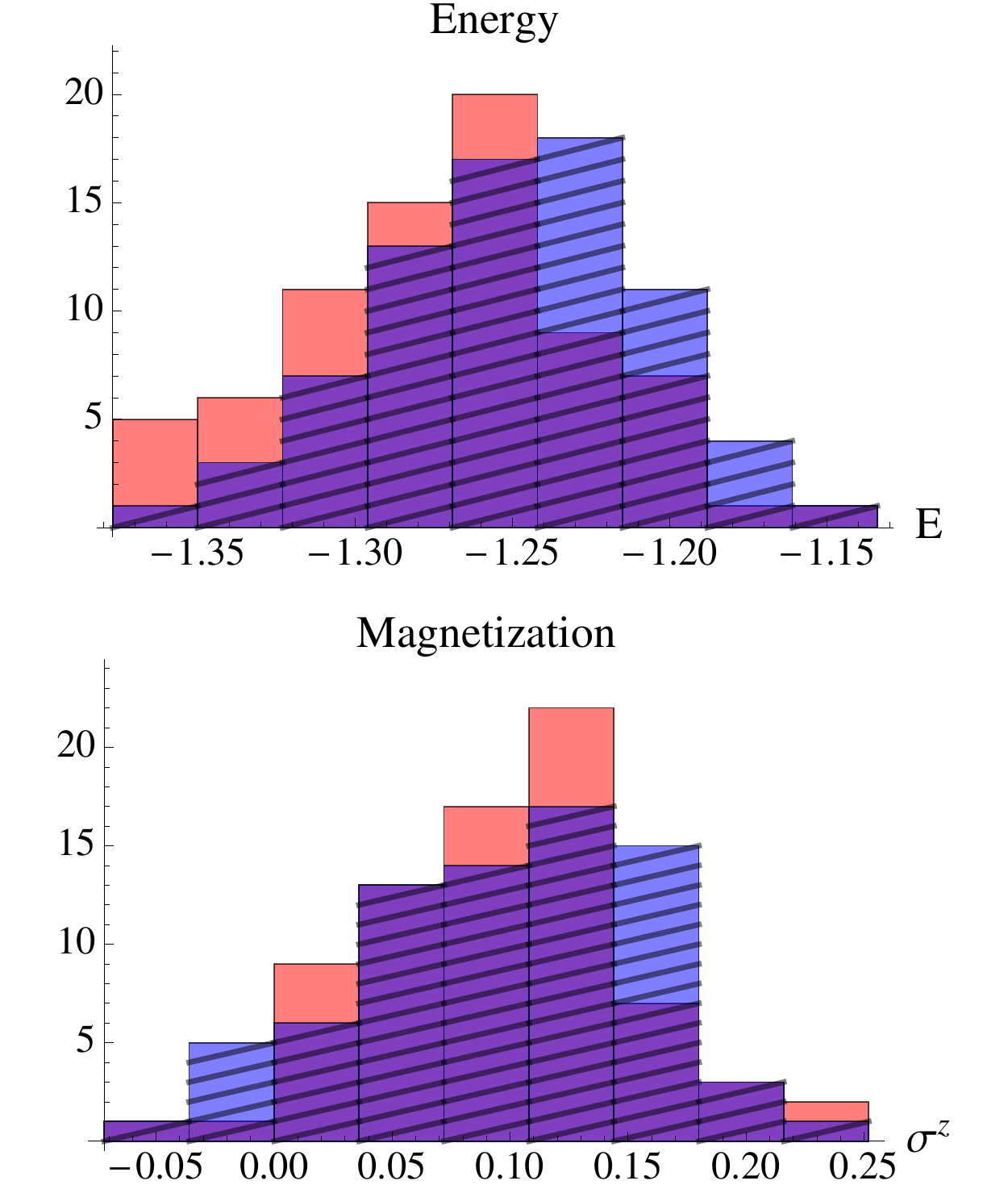}
\caption{(Color online) Benchmarking the flow equation method (red bins) against simulated annealing (blue, hatched bins) for the Ising spin glass problem in Eq.~\eqref{Hising}. Shown are histograms of ground state energy per lattice site $E/N$ and magnetization $\langle \sigma^z \rangle$ for 100 instances of random $J_{ij}$'s, obtained after minimizing \eqref{ising} for random Gaussian nearest neighbor $J_{ij}$'s with zero mean and unit variance, $h_x=0.05$ and $h_z=0.1$ on a square lattice with $15 \times 15$ sites and periodic boundary conditions. Energies are in units of $\sqrt{\text{Var}(J_{ij})}=1$.}
\label{fig:hist}
\end{center}
\end{figure}

Now we're going to apply the flow equation to find variational ground states of frustrated Ising models with random- or long-range antiferromagnetic interactions in a combined longitudinal- and transverse field on a two dimensional square lattice. The Hamiltonian has the form
\begin{equation}
H = \sum_{ i<j } J_{ij} \, \sigma^z_i \sigma^z_j - h_x \sum_i \sigma^x_i - h_z \sum_i \sigma^z_i \ ,
\label{Hising}
\end{equation}
where $\sigma^z_i$ and $\sigma^x_i$ are Pauli matrices on lattice site $i$ , and the exchange couplings $J_{ij}$ are either restricted to nearest neighbors and randomly drawn from a Gaussian distribution with zero mean and unit variance, or decrease with a power law $J_{ij} = 1/|\mathbf{r}_i -\mathbf{r}_j|^\alpha$ as a function of distance between the two sites, where $\mathbf{r}_i$ denotes the position of lattice site $i$. 
As variational ansatz for the ground state we use the most general product state 
\begin{equation}
|\psi_0 \rangle = \prod_i \big( \alpha_i  | \! \uparrow_i \rangle + \beta_i | \! \downarrow_i \rangle \big) \ ,
\end{equation}
where $ | \! \uparrow \rangle$ and $ | \! \downarrow \rangle$ are the two eigenstates of $\sigma^z$. Note that this ansatz gives the exact ground-state in the classical Ising limit $h_x \to 0$, but doesn't include quantum correlations between the spins.
%Our task is thus a minimization of the ground-state energy with respect to the variational parameters $\alpha_i$ and $\beta_i$, where $i \in \{ 1, \dots, N\}$ and $N$ denotes the number of lattice sites. 
In the ground-state the parameters $\alpha_i$ and $\beta_i$ can be chosen to be real. Using the normalisation condition $\alpha_i^2+\beta_i^2 = 1$ we reduce the problem to a minimization of the energy with respect to probabilities $f_i := \beta_i^2 \in [0,1]$ to be in the spin-up state 
\begin{eqnarray}
E[f_n] &=&  \sum_{i<j} J_{ij} (2 f_i  - 1) (2 f_j  - 1) - h_z \sum_i (2 f_i - 1) \notag \\
&& - 2 h_x \sum_i \sqrt{f_i (1-f_i)} \ .
\label{ising}
\end{eqnarray}
In the following we compare results of minimizing \eqref{ising} with the flow equation method, simulated annealing and other gradient based methods. 

\subsection{Spin glasses}
\label{sec4a}

As a first example we choose random nearest-neighbor $J_{ij}$'s from a Gaussian distribution with zero mean and unit variance. In this case Eq.~\eqref{Hising} represents an Edwards-Anderson model in a combined longitudinal- and transverse-field, which is a prime example of a model for spin-glasses \cite{Edwards}. We minimize \eqref{ising} on a square lattice with $15 \times 15$ sites and periodic boundary conditions, using the flow equation method as well as simulated annealing. In Fig.~\ref{fig:hist} we show histograms of ground-state energy and magnetisation in z-direction $\langle \sigma^z \rangle = \sum_i \langle \sigma^z_i \rangle /N$, obtained by minimizing $100$ different random instances, starting from random initial states with average magnetisation close to $\sigma^z_i \simeq 1/2$. For each instance we use 10 different initial conditions and choose the solution with the lowest energy.
Shown are results for $h_x=0.05$ and $h_z=0.1$. Average energies, magnetizations and their variances are listed in table \ref{tab1}. The results are comparable, even though the flow equation gives slightly lower energies. Its main advantage, however, is a significant speed-up in the simulation time by a factor $4.8$ as compared to simulated annealing. Note that Eq.~\eqref{floweqdiscrete} was solved using a two-stage Runge-Kutta scheme (Heun's method) with adaptive step size. Simulated annealing was performed using the GSL library \cite{GSL}, where the temperature was lowered from $T=0.01$ to $T=5 \cdot 10^{-5}$ using a damping factor $1.05$ with $10^4$ iterations at each temperature and a maximum step size of $0.01$.

\begin{table}
\caption{Comparison between the flow equation method and simulated annealing for the Ising spin glass problem defined in Sec.~\ref{sec4a}. Listed are the mean energy $E/N$ and magnetisation $\sigma^z$ corresponding to the histograms shown in Fig.~\ref{fig:hist}, together with their variances.}
\begin{ruledtabular}
\begin{tabular}{c| c  c  c  c }
& $E/N$ & $\sqrt{\text{Var}(E)}/N$ & $\sigma^z$  &  $\sqrt{\text{Var}(\sigma^z)}$  \\
\hline
flow eq. & -1.272 & 0.046 & 0.097 & 0.054  \\
sim. ann. & -1.249 & 0.043 & 0.100 & 0.061 
\end{tabular}
\end{ruledtabular}
\label{tab1}
\end{table}

\subsection{Frustrated Ising models with dipolar interactions}
\label{sec3b}

\begin{figure}
\begin{center}
\includegraphics[width=0.95\columnwidth]{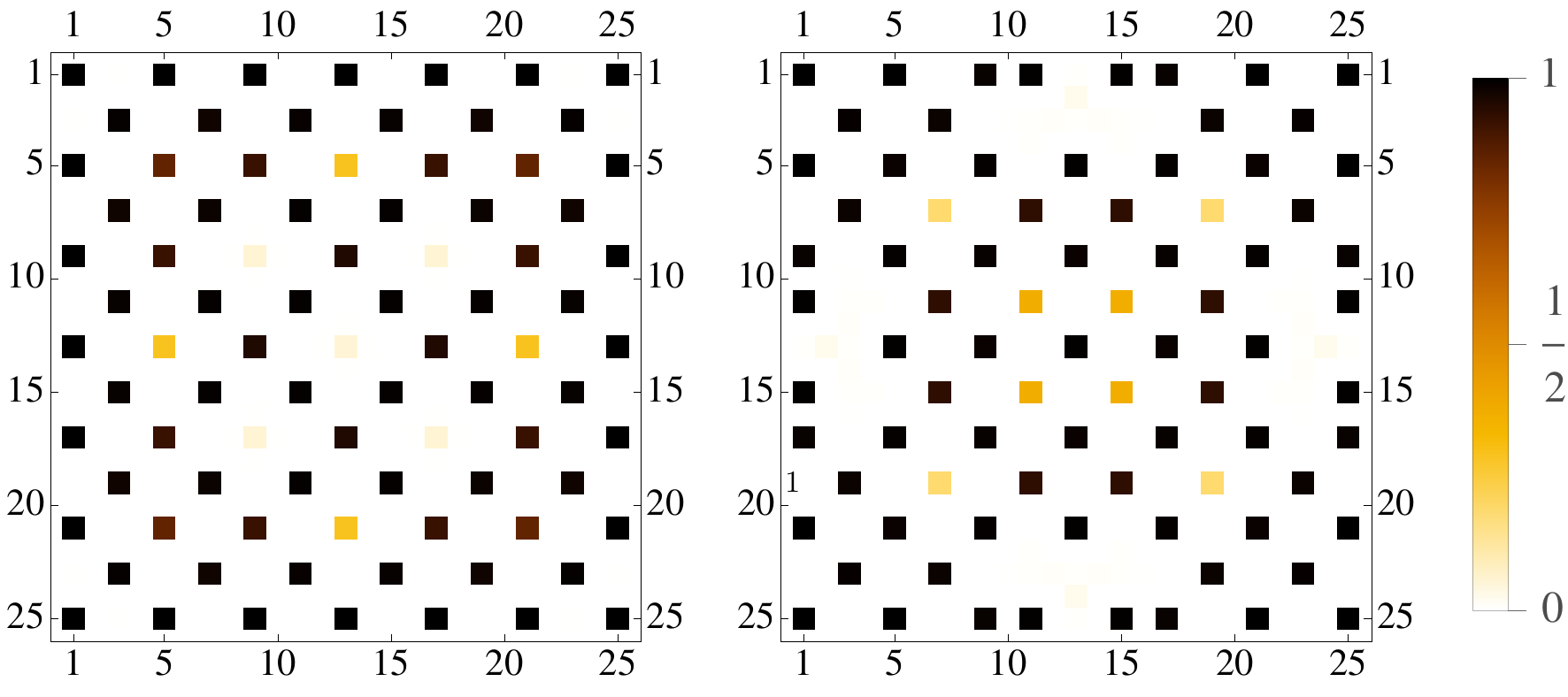}
\caption{(Color online) Benchmarking the flow equation method (left) against a quasi-Newton BFGS algorithm (right) for the Ising model in \eqref{Hising} with dipolar interactions $J_{ij} = 1/|\mathbf{r}_i - \mathbf{r}_j|^3$, slightly above the saturation field where all spins point down. Plotted is the probability $f_i = \beta_i^2$ to find a spin pointing up on a given site of a $25 \times 25$ lattice with open boundary conditions. The probabilities were obtained by minimizing Eq. \eqref{ising} with dipolar interactions for $h_x=0.02$ and $\tilde{h}_z = h_z + \sum_j J_{ij} = 0.6 $ (see text). The corresponding energies are $E/N = -4.00686$ (left) and $E/N = -4.00474$ (right). Energies are measured in units of the nearest-neighbor exchange coupling $J_{i,i+1}=1$.}
\label{fig:dipolar}
\end{center}
\end{figure}

Finally, in Fig.~\ref{fig:dipolar} we show results of minimizing Eq.~\eqref{ising} with dipolar interactions $J_{ij} = 1/|\mathbf{r}_i - \mathbf{r}_j|^3$ on a square lattice of $N = 25 \times 25$ sites with open boundary conditions at large negative longitudinal fields $h_z = \tilde{h}_z - \sum_j J_{ij}$ slightly above the saturation field where all spins point down. Shown are results for $h_x = 0.02$ and $\tilde{h}_z = 0.6$. Finding the ground-state in the regime close to the saturation field is particularly hard, because in the classical limit $h_x \to 0$ this corresponds to finding the minimum energy configuration of classical dipoles on a lattice \cite{Rademaker}. The up-spins want to maximize their distance and form a triangular lattice, which is not possible due to the underlying square lattice. Incommensurability issues thus play an important role, resulting in many local energy minima.

In Fig.~\ref{fig:dipolar} we compare the flow equation method with a quasi-Newton BFGS algorithm \cite{BFGS}. For the results shown here a continuous update of the prior was not performed, because the convergence turned out to be slightly better. In both cases we initialize all $f_i = const.$ and find that our flow equation leads to a state with energy $E/N = -4.00686$, which is always lower than the state obtained using the BFGS algorithm, where the lowest energy state we found has $E/N = -4.00474$ (energies are measured in units of the nearest-neighbor exchange coupling $J_{i,i+1}$). Moreover, the results of the BFGS algorithm are extremely sensitive to the initial condition and one easily ends up in a local minimum with high energy, whereas our flow equation is rather insensitive to the choice of initial conditions. As can be seen in Fig.~\ref{fig:dipolar}, the up-spins try to maximize their distance and form a regular arrangement which is pinned by boundary effects. We also tried simulated annealing to find the variational ground-state, but the results were rather poor with energies around $E/N = -3.974$. 

Note that the model in Eq.~\eqref{ising} with dipolar interactions on a square lattice with open boundary conditions is symmetric under rotations by 90 degrees, as well as inversions about the x-, y-axis and the diagonals. The flow equation \eqref{floweqdiscrete} only depends on derivatives of Eq.~\eqref{ising} and thus respects the symmetries of the problem. Accordingly it preserves the symmetries during the flow, if the initial condition is symmetric as well. For the problem at hand it is not known wether the ground-state breaks these symmetries, but the symmetric solutions we found were always lower in energy than solutions found by starting from a random initial condition.

\section{Conclusions}

We presented a multidimensional minimization method which combines ideas of homotopy continuation and the maximum entropy principle and applied it successfully to problems in condensed matter physics. We note that this flow equation method has some similarities with quantum annealing \cite{Finnila, Kadowaki} or adiabatic quantum computation \cite{Fahri}. In our approach we adiabatically deform a product state to determine the ground state of a classical Ising model, rather than following the exact ground-state of the full quantum model adiabatically to the Ising limit, starting from large $h_x$ (for a similar idea see also \cite{Hsu}). 

Finally we note that it is straightforward to generalize the flow equation \eqref{floweq} to a stochastic differential equation by adding a random force term which might be able to kick the flow out of a local minimum. By properly choosing a random force term which vanishes sufficiently fast for $t \to 1$ it might possible to construct a method which has similarities to simulated annealing and which has an even higher success rate finding the global minimum. We leave such extensions of this method open for further investigation.

\begin{acknowledgements}

We thank A.M. L\"auchli and W. Lechner for discussions and W. Zwerger for comments on the manuscript. This work was supported by the Erwin-Schr\"odinger Fellowship J 3077-N16 of the Austrian Science Fund (FWF).

\end{acknowledgements}

\end{document}